\documentclass[12pt]{spieman}

\usepackage{amsmath,amsfonts,amssymb}
\usepackage{graphicx}
\usepackage[colorlinks=true, allcolors=blue]{hyperref}
\usepackage{setspace}
\usepackage{tocloft}
\usepackage{lineno}
\usepackage{comment}
\usepackage{color,soul}

\title{Ground-Based Astronomical Instrumentation Development in the United States:  A White Paper on the Challenges Faced by the US Community}

\author[a,*]{Stephen A. Smee}
\author[b,c]{Gary J. Hill}
\affil[a]{Johns Hopkins University, Department of Physics and Astronomy, 3701 San Martin Drive, Baltimore, MD 21218, USA}
\affil[b]{McDonald Observatory, University of Texas at Austin, 2515 Speedway, Stop C1402,  Austin, TX 78712, USA}
\affil[c]{Department of Astronomy, University of Texas at Austin, 2515 Speedway, Stop C1400, Austin, TX 78712, USA}

\cftpagenumbersoff{figure}
\cftpagenumbersoff{table} 
\begin{document} 
\maketitle

\begin{abstract}
This invited white paper, submitted to the National Science Foundation in January of 2020\footnote{Original submission date: January 22, 2020.} discusses the current challenges faced by the United States astronomical instrumentation community in the era of extremely large telescopes (ELTs).  Some details may have changed since submission, but the basic tenets are still very much valid.  The paper summarizes the technical, funding, and personnel challenges the US community faces, provides an informal census of current instrumentation groups in the US, and compares the state-of-affairs in the US with that of the European community, which builds astronomical instruments from consortia of large hard-money funded instrument centers in a coordinated fashion.  With the recent release of the Decadal Survey on Astronomy and Astrophysics 2020 (Astro2020), it is clear that strong community support exists for this next generation of large telescopes in the US.  Is the US ready?  Are there sufficient talented instrumentalists, facilities, and resources in the community today to meet the challenge of developing the complex suite of instruments envisioned for two US ELTs?  These questions are addressed, along with thoughts on how the National Science Foundation can help build a more viable and stable instrumentation program in the US.  These thoughts are intended to serve as a starting point for a broader discussion, with the end goal being a plan that puts the US astronomical instrumentation community on solid footing and poised to take on the challenges presented by the ambitious goals we have set in the era of ELTs. We provide an epilogue that includes reference to the 2020 Decadal Survey and an update of some details that have changed since the original white paper was submitted.
\end{abstract}


{\noindent \footnotesize\textbf{*}Stephen A. Smee,  \linkable{smee@jhu.edu} }

\begin{spacing}{1}   

\section{EXECUTIVE SUMMARY}
\label{sec:exec_sum}  
Astronomy as a field would not exist as we know it without continual instrumentation development.  For centuries the field has advanced in lock-step with the development of ever-more-advanced instruments fed by larger and larger telescopes.  And these advancements have required increasingly sophisticated technology, time and money to develop.  Long gone are the days when an astronomer like Galileo could craft a simple telescope in his workshop and change the world with its observations. Modern ground-based optical telescopes require instruments that take of order a decade to develop, consuming tens of thousands of hours of engineering and millions of dollars in the process.

Who builds these instruments? Scientists, engineers, and craftsmen do. In Europe, astronomical instruments are built by a collective of hard-money funded university research groups and research centers.  In the United States, instruments are built largely by small, mostly soft-money funded, university research groups. The Europeans have the advantage of steady funding and better coordination, which leads to permanent, stable, technical staff with the specialized skills needed for this type of work.  In the United States the situation is quite different. The lack of hard-money funding leads to a sparse network of small university research groups, very few of which ever grow to sufficient size to build a modern instrument without substantial aid from external collaborators, be it other universities or commercial entities.  In short, the lack of hard-money funding in the US has led to a situation where very few institutions have the breadth and depth of expertise to construct a major instrument.  And with two US-led thirty meter telescopes on the horizon, and a network of 8-meter class telescopes currently in operation and in need of upgrades, the United States is woefully short of the technical talent necessary to satisfy the ambitions of the US ground-based astronomical community.  We have a people problem.

We have a technology problem as well.  In some key technology areas we lack sufficient depth in the vendor pool. In some cases instrument builders are limited to even a single qualified vendor to provide a key component or service. In other cases, consolidation in the market threatens to reduce competition altogether, driving up costs.  At some point, we as a community must ask ourselves if there is a way to better control our destiny.

We need to fix these problems, and the National Science Foundation (NSF) can help, by providing hard-money funding to establish a small network of Centers of Excellence (CoE) in astronomical instrumentation throughout the United States. These centers would fall into one of two categories: Engineering centers and Technology centers.  Engineering centers would serve as a highly skilled technical team having the size and core technical competence to construct modern instruments. These centers would receive strategic funding to develop robust instrument concepts for National observatories, and then build those instruments through a competitive process.  Projects would be carried out in a more efficient way, with fewer external interfaces and clear lines of responsibility, ultimately leading to instruments of the highest possible performance and reliability. Technology centers would focus on advancing the key technologies needed for modern instruments, and transitioning that technology to the community.  Collectively, these centers would serve as a training ground for students and post-docs, providing a sustained and highly talented workforce, well positioned to develop the next generation of instrumentation.

\section{INTRODUCTION}
\label{sec:intro}  

Over the years scientific instrumentation has become increasingly complex, and ground-based astronomical instrumentation is no exception.  Even in the past twenty-five years the landscape has changed significantly.  Instruments built today for a modern 8~m class telescope take a minimum of six years to develop and cost between approximately \$10M and \$100M.  They are significant engineering endeavors consuming, in many cases, well over a hundred  FTE (Full Time Equivalent) years to complete. The thirty-meter telescope instruments will be even more complex, expensive, and labor intensive.  It is doubtful that anyone really knows how challenging it will be to build this next generation of instruments, but as a point of reference, the current manpower estimate to build HARMONI\cite{2021Msngr.182....7T}, the workhorse integral field spectrograph for the European Extremely Large Telescope (E-ELT)\cite{2016SPIE.9906E..0WT}, is approximately 600 FTE-years. Labor estimates for developing facility class instruments for the Giant Magellan Telescope (GMT)\cite{2020SPIE11445E..1FF} and the Thirty Meter Telescope (TMT)\cite{2014tmt..confE..60S} are comparable.  Of course, it is too early to say what the real numbers are, and it will take at least a decade to find out.  Suffice it to say, a typical facility class instrument for the E-ELT, TMT, or GMT will take hundreds of FTE-years to develop over a period of ten years or more. Moreover, the expectation is to deploy a new instrument for each ELT every 5 years. 

Where will this technical talent come from? 

In Europe, major astronomical instruments are built by a collective of well-established, mostly-soft-money funded university research groups and hard-money funded research centers having very experienced, seasoned professional engineers.  The Europeans have the advantage of steady, predictable funding profiles and better coordination, which leads to permanent, stable, technical staff with the specialized skills needed for this type of work.  
This deep pool of expertise is applied to large projects through the formation of consortia of institutions. The labor is the currency for their involvement.

In the United States the situation is quite different. Instruments are built largely by small Principal-Investigator-led groups consisting of a handful of technical staff, often relying on students and post-docs to round out the labor force.  These technical staff positions are often soft-money funded.  When the project is finished the technical staff move on.  In some cases they find a position at the same institution, or they find a position somewhere else, possibly in another field. In short, the situation is tenuous at best.  At present, there are of order a dozen groups building optical instruments for ground-based telescopes in the US. Most have fewer than ten non-faculty full-time professional staff.  Most do not possess all the expertise needed to build a modern 8~m class instrument without some form of collaboration. And most do not have the facilities either.

It has been our experience over the years that success or failure in these endeavors hinges on people - their skill, creativity, experience, motivation, etc.  Having the right people, and enough of them, is the difference.  Today we have two US-led thirty meter telescopes on the horizon and a network of 8~m class telescopes currently in operation and in need of upgrades.  One does not have to examine the situation very hard to realize that the United States is woefully short of the technical talent needed to satisfy the needs of the US ground-based astronomical community.  In short, we have a people problem.

We have technology problems as well.  In some key technology areas we lack sufficient depth in the vendor pool. In some cases instrument builders are limited to a single qualified vendor to provide a key component or service. In other cases, consolidation in the market threatens to reduce competition altogether, driving up cost.  At some point, we as a community must ask ourselves if there is a way to better control our destiny.  

Collectively, there are numerous challenges facing the US ground-based instrument community, which can be distilled down to a set of fundamental questions:  Where will all the technical talent come from to build these new instruments in the era of 30~m class telescopes?  How do we sustain/retain top-tier instrumentation talent within the US when these folks operate in a soft money world, and when steady funding to instrument groups is so tenuous? How do we grow instrument groups to reach critical mass, whereby sufficient technical expertise exists within an institution to develop 8~m and 30~m class instruments.  How do we bring more young people into the field and provide a sustainable career path for them? 
How do we increase involvement from underrepresented groups? 
How do we address the issue of limited vendors in key technology areas?  How do we develop and maintain key instrument technologies that are critical to advancing the state of the art?

This white paper discusses the current state of affairs with regard to the development of optical instrumentation for ground-based telescopes in the United Sates; in particular, the challenges we face as a community, thoughts on how we can do better, and how the National Science Foundation can help. 

\section{The United States vs. Europe}
\label{sec:US_vs_ESO}

The development of ground-based telescope instruments in the United States and Europe is similar in some ways and different in others. Here we examine the differences to understand where the relative strengths and weaknesses are.

\subsection{The European Model}
\label{sec:ESO}

The path to developing a major ground-based optical instrument in Europe starts, in general, at ESO (European Southern Observatory), which is funded by annual contributions from the 16 Member States. Science workshops are held to solicit ideas from the community for future instrumentation.  From that dialogue, a committee of ESO members works to define the instrument desired, from which a call for proposals is released to the community.  Proposals are submitted by interested teams comprised of a lead investigator and one or more science and instrumentation centers throughout Europe that will provide the technical talent to carry out the work.  These instrumentation centers, located primarily in Germany, France, the United Kingdom, Italy, and Holland, are state or regional government funded institutions that employee full-time technical staff with the skills and experience necessary to build instruments, and facilities to match.  Once a proposing consortium is selected, that team builds the instrument.  Funding for the hardware comes from ESO, i.e. from the member countries.  Labor is provided by the instrumentation centers with Guaranteed Time for Observing (GTO) going to the centers providing the labor in proportion to the contributed effort.  It should be said that aside from the ESO telescopes ESO Member States often have additional telescope access through partnerships in non-ESO observatories; e.g. CFHT (Canada-France-Hawaii Telescope) for France, LBT (Large Binocular Telescope) for Germany, GranTecan (Gran Telescopio CANARIAS) for Spain.  Instrumentation centers in these countries contribute to instruments for these observatories as well.

These instrumentation centers are key to the process, and the success of ESO instruments.  They provide the specialized technical talent required to build modern instruments.  They range in size from small groups of order ten full-time technical staff to larger groups of order 50 full-time staff.  These government-funded centers provide a stable workforce, retaining experienced talent that takes years, if not decades, to acquire.  They represent a commitment by the member countries to a scientific endeavor that only advances when technology pushes it.  A commitment that will pay off in the era of extremely large telescopes where the instruments are very complex, requiring large teams of highly specialized technical staff, many years, and significant funding to develop.

\subsection{The Model in the United States}
\label{sec:US}
Developing ground-based optical instrumentation in the United States is, on the surface, similar to the European model. Ideas for new instruments stem from community input in the way of workshops, meetings, and the Decadal Survey\cite{Astro2010}. From there, the process diverges a bit depending on the funding source.  For observatories relying on private funding to build an instrument, a lead investigator secures the funds, builds a technical team and carries out the work. For instrumentation funded by the National Science Foundation, ideas for instruments are competed, there are calls for proposals, lead investigators submit proposals, and awards are granted based on a combination of criteria that considers the strength of the team, the impact of the science, and of course the budget.

However, there is a difference between Europe and the US, and it matters.  In the United States, there are no nationally funded centers for developing instruments. Instruments have historically been developed by technical teams at the the university where the lead investigator resides. The teams are small ($\sim$ 5 full-time staff) compared to those at the technology centers in Europe, and are often not large enough to develop an instrument without external collaboration. They lack the full complement of engineering skills  and facilities required to build a modern instrument, relying too heavily on students to round out the workforce.  And they have historically been soft-money funded.  Funding beyond the project most often does not exist to retain professional talent long term, nor to justify the investment in the equipment and facilities needed to build state-of-the-art instruments.  Hence, these groups are, \emph{by design}, small and ill-equipped compared to our European counterparts. In short, most university groups do not have the resources to build a 30~m class telescope instrument.

\section{Building a Modern Optical Instrument}

What is required to develop a modern ground-based instrument?  Time, money and people. More specifically, \emph{enough} time, \emph{enough} money and the \emph{right} people; scientists and engineers, but these days mostly engineers, as these projects have increasingly required professional and highly skilled engineers to carry out.

\subsection{How long does it take?}
A new optical instrument for an 8~m class optical telescope takes, realistically, a minimum of six years to develop, from the start of conceptual design to commissioning. Of course, schedule depends on the complexity of the instrument and eight to ten years is not unheard of. 

The time required to develop a 30~m class optical telescope instrument, i.e. GMT, TMT, E-ELT, is not known, but a minimum of ten years is a reasonable expectation. The moderate resolution optical spectrograph, GMACS\cite{2018SPIE10702E..1XD}, for GMT is expected to take a minimum of ten years to develop. Both HARMONI on the E-ELT, and G-Clef\cite{2018SPIE10702E..1RS}, HARMONI's counterpart on GMT, are expected to take 20 years to develop.  

\subsection{How much money is required?}
 A new instrument for an 8~m class telescope costs of order \$10M to \$100M, including labor, depending significantly on the complexity of the instrument. An instrument like SCORPIO\cite{2018SPIE10702E..0IR}, a broadband optical/near-IR long-slit spectrograph being developed for Gemini-South, will fall near the lower end of that range.  The Prime Focus Spectrograph (PFS)\cite{2016ASPC..507..387T} for the Subaru Telescope, a highly multi-plexed optical/near-IR fiber spectrograph is pushing the upper end of this range. For the US 30~m class telescopes, current rough order of magnitude estimates range from \$30M to \$100M.

\subsection{What Skills are Required?}
Virtually all instruments require optical, mechanical, electronics, detector system, and software engineers.  Delving deeper, the optical engineers need to be well versed in lens design, optical system design, stray light, ghosting, coatings, alignment and testing.  The mechanical engineers needs to be well versed in optomechanical, structural, thermal, and vacuum system design, have a good understanding of manufacturing methods, and have solid analysis skills (thermal, structural, vibration).   Electronics engineers require digital and analog system skills, controls, and preferably board-level analog and digital design skills for the development of custom boards.  Detector engineers need to understand how to characterize detectors and optimize their performance, as well as a host of the skills just mentioned to deal with the test infrastructure; these are all specialized skills.  Lastly, software engineers need to be well-versed in coding for system control, instrument operation, and data storage; then there is the pipeline software, which is its own specialty.  And, of course, all these engineers need to be well-versed in the nuances of the field, knowing what works and why.

There are several other key rolls we should not forget.  One is the lead investigator, who typically has overall responsibility for the design and construction of the instrument.  In the US this is often an astronomer that not only knows the science but has a fair bit of engineering knowledge as well.  That engineering knowledge, while not necessarily essential, is very valuable.  It allows him/her to understand the engineering details well enough to guide the decision making process on technical matters if needed.  But the most important roll for the lead investigator, aside from the science itself, is to ensure that the engineering requirements satisfy the science requirements.

The second is the instrument scientist.  This position is in some respects similar to the lead investigator but with greater emphasis on providing day-to-day support for various engineering activities.  This person manages the transition from science requirements to engineering requirements, working closely with the systems engineer and engineering leads.  

The third is the project manager.  The project manager has overall management responsibility for the technical workforce developing the instrument.  In most cases, this needs to be a seasoned individual familiar with all phases of instrument development, the strengths/weaknesses of the personnel at his/her disposal, and likewise, the strengths/weaknesses of the vendor pool the project will be relying on for outsourced hardware.  The project manager manages the schedule and budget and must employ sound judgement in the decision making process with regard to both.

A fourth roll, which comes into play strongly on large projects where multiple institutions are involved, is the systems engineer.  The systems engineer manages sub-system interfaces and instrument requirements, working to ensure that individual sub-systems are compliant at the instrument level and the overall instrument performance requirements are met.  This person often manages documentation as well, ensuring that the design and testing of the instrument are properly documented.

Lastly, but certainly not least, are the fabricators (i.e. machinists, welders etc.) that build the various components.  It cannot be overstated how important good machinists are in this business.  Mounting and aligning optics is a precision endeavor and the mounts for these components most often need to be made very precisely.  
Some fabrication can be outsourced, but experience shows that there is no substitute for highly skilled in-house machinists for critical items.
Finally, integration and test requires  technicians that help to put it all together and test it.  

In total, it is an extensive list. One that, at best, a handful of universities can cover with the staff they have.  There are many universities with subsets of the required expertise, and where they fall short they outsource or make do.

\subsection{Why are Universities Good Places to Build Instruments?}
\label{sec:why_U}
Universities have historically been the place where instruments are developed, in large part because universities are where the lead investigators are, and new ideas are often spurred in this environment.  The tight coupling between the scientist leading the effort and the technical staff working the details has obvious advantages.  But there are other reasons why universities are a good place to build instruments. 

First, they are cost effective.  Universities often reduce or eliminate overhead when State or private funding is involved. Universities do this because they have a vested interest in the science, and reducing or eliminating overhead is a way to do more with available funding.   Doing more with less is necessary in a funding-limited environment such as ground-based astronomy instrumentation.  It should be said that the general community leverages this overhead relief when public access is granted through National Science Foundation buy-in.

Second, universities are magnets for highly motivated and creative people; the type of people needed to build complex instruments with limited budgets and aggressive schedules.  Large matrixed organizations can be bureaucratic, often requiring requisition and approval to do even the simplest things, such as purchasing a box of screws.  Motivated people will struggle in such an environment.  In addition, matrixed organizations often confine employees to technical lanes, or areas of specialty, rather than providing the freedom to branch out into other disciplines. Highly motivated, multi-disciplined scientists and engineers are what is needed in a funding-limited environment. Fewer people, fewer emails, fewer meetings, less management, fewer interfaces to manage, less systems engineering, and more progress per unit time. It should be said that universities are not the only place where such environments exist, but rather universities are more likely to be a place where one would find such an environment.

Third, innovation comes naturally from the university environment, and innovation is needed to drive the state of the art. 

Finally, universities are where bright students are, the next generation of instrument builders.  And we need to train them to do the next big thing.

\section{Current Instrument Groups in the United States}
\label{sec:groups}

\subsection{University Groups}
In the United States, there are about a dozen university instrument groups of varying size.  The vast majority are small, with less than ten technical non-student personnel (i.e. engineers and/or scientists) working full-time (or majority-time) on the development of instrumentation. Groups have come and gone over the years, popping up or inflating in size with fresh instrument projects, then downsizing, or in some cases disappearing, when projects no longer exist.  Today what remains is a mix of hard-money and soft-money funded groups.  But more and more, soft-money groups are disappearing.

Table~\ref{tab:census} provides a list of instrument groups in the US that responded to our informal census, or those for which reliable information was available from an alternate source. Listed are the observatories they support and the level of hard-money and soft-money full-time professional staff: technical, managerial, and dedicated administrative personnel.  Faculty are not included in this calculus.  Note that the subject of supporting observatories is not straight-forward.  In some cases the support is direct.  In others it is indirect via institutional affiliation with the entity that funds the observatory.

Here, hard-money positions are those positions funded by sources that are, essentially, permanent; for example, positions where salary is derived from State Government funds or university endowment. Soft-money positions are those positions funded by temporary sources of income, such as grants, non-endowed private donations, university research funds that are temporary, or funding from external sources for a fixed level of effort.

Table 1 does not list several smaller programs with only a couple of full-time technical staff (e.g. University of Washington, University of Wisconsin, Madison, and others). 

Table 1 also does not list the University of Arizona/Steward Observatory, which has a large staff of roughly 60, supporting several major optical/IR telescopes in Arizona as part of Steward Observatory (LBT, MMT, Mt Lemmon, Kitt Peak); we were not able to obtain numbers from them in time for this paper.  Many of the technical positions are funded by the university and are effectively hard money, with the majority supporting telescope operations rather than new instrument development. There are no hard money positions dedicated to ground-based instrument development.  The large pool of talent available at UA/Steward can be drawn upon for instruments, usually projects that are funded by external grants.  In addition, the Imaging Technology Lab (ITL) provides detectors and readout systems, while being primarily supported by external contracts.
The significant adaptive optics program of the Center for Astronomical Adaptive Optics (CAAO) also has an instrumentation component. Recent science instruments include the LBT Interferometer\cite{2003SPIE.4838..108H}.


From this census, the four largest university groups in the US building instruments for optical/infrared ground-based telescopes are at Caltech, the University of Texas at Austin (McDonald Observatory), the University of California Santa Cruz (UCSC), and the Johns Hopkins University (JHU); Caltech's being the largest at approximately 26.5 full-time personnel. UT Austin has 20.5 full-time staff. UCSC has 15 full-time staff.  And the Johns Hopkins University has 14 full-time staff.  

The Optical/Infrared Instrumentation Group within the Astronomy Department at Caltech is dedicated to optical and infrared ground-based instrumentation and develops instruments primarily for the TMT, Keck, and Palomar Observatories.  Funding for the group comes largely in the form of soft-money, with some hard-money support from the observatories if needed during times of limited work.  The group has very good facilities and an excellent detector lab. 
It also benefits from close association with the Jet Propulsion Lab (JPL). Caltech also has separate groups dedicated to instrumentation for ultraviolet, radio and submillimeter astronomy, as well as for space observatories.  

The McDonald Observatory (MDO) instrument group at the University of Texas at Austin develops a range of instrumentation for optical and near IR astronomy with an emphasis on high resolution spectroscopy (optical and NIR), and fiber-fed moderate resolution integral field spectrographs. The instrument development group in Austin has four Research Professor/Scientist PIs, five mechanical engineers, two software engineers, two technicians, and four machinists on the permanent staff plus several people working on soft money. Most instrument development is aimed at MDO telescopes or the Hobby-Eberly Telescope, but instruments have been supplied to the Discovery Channel Telescope (DCT) and the Gemini Observatory, and new development for the Giant Magellan Telescope is underway. MDO has developed particular technical expertise in mass-production, immersed Silicon gratings, fiber optics, and optical testing.

The University of California Santa Cruz instrument group is home to one of two technical facilities under the University of California Observatories (UCO) umbrella.  The second technical facility is the instrument group at UCLA.  UCO operates Lick Observatory, is a managing partner for the Keck Observatory, and is the UC center of participation for the Thirty Meter Telescope.   The UCSC technical facilities include an engineering department, an electronics lab, an optical lab, an instrument lab, and an optical coatings lab. The UCSC group builds most of the instruments for Lick Observatory, and has built some of the Keck instruments, including the Deep Imaging Multi-Object Spectrograph (DEIMOS)\cite{1997SPIE.2871.1107C}, The High Resolution Echelle Spectrometer (HIRES)\cite{1994SPIE.2198..362V}, and the Echellette Spectrograph and Imager (ESI)\cite{2002PASP..114..851S}. Current projects include the development of a low resolution spectrometer for targets of opportunity (DARTS) for the Automated Planet Finder (APF)\cite{2014PASP..126..359V} telescope on Mt. Hamilton. The group is also working on instrument concepts for future Keck instruments:   an exoplanet AO imager called SCALES\cite{2021LPI....52.2412S}, and a multi-object fiber-based spectrometer called FOBOS\cite{2019BAAS...51g.198B}.  The lab employs two scientists, four and a half engineers, one and a half project managers, one software/pipeline developer, four technicians/machinists, and the equivalent of two full-time administrators.  All are on soft money.

The Instrument Development Group (IDG) in the Department of Physics and Astronomy at Johns Hopkins University develops instrumentation for physics and astronomy research, but mostly for astronomy; a combination of UV/optical/infrared ground-based instruments as well as instrumentation in support of space observatories. The Department of Physics and Astronomy has a long-standing history of instrument development, dating back to the late 1800s, and similar to Caltech, has additional research groups dedicated to developing astronomy instrumentation; one dedicated to the far-Ultraviolet from space, and the other dedicated to the microwave region of the electromagnetic spectrum.  Like Caltech, UCSC, and UT Austin, the IDG is comprised almost exclusively of professional engineers, each having expertise in one of the core disciplines related to this field: optical, mechanical, electronics, and software. The group is unique amongst its peers in that it has a state-of-the-art machine shop, equipped specifically for research instrumentation.  The group has operated entirely on soft money since its inception.

\begin{table}

\centering
\caption{Census of University instrument groups, support model, and staff levels:}
\vspace{3mm}
\begin{tabular}{l|l|c|c|c}
\label{tab:census}

    University & Supporting Observatories &  Hard Money &  Soft Money & Total\\
     & & Staff & Staff & Staff\\
    \hline
    Caltech & Palomar, Keck, TMT &  & 26.5  & 26.5\\
    UT Austin & McDonald, HET, GMT & 17.5 & 3 & 20.5\\
    UCSC & Lick, TMT &  & 15 & 15\\
    JHU & APO &  & 14 & 14\\
    UCLA & Lick, Keck, TMT & 4 & 4 & 8\\
    OSU & LBT & 7.5 &  & 7.5\\
    MIT & Magellan &  & 7 & 7\\
    U. Florida & GTC & 4 & 1 & 5\\
    UVA &  & 4 & 1 & 5\\
    Penn State & HET, WIYN &  & 4 & 4\\
    Texas A\&M &  GMT & 1 & 3 & 4 \\
    \hline
    Totals & & 38 & 78.5 & 116.5\\
    \hline

\end{tabular}
\end{table}

\subsection{Research Institutes}
In addition to these university groups, there are two non-university instrument groups that are noteworthy.  One, is a collection of personnel at Carnegie Observatories, who dedicate most/all of their time to developing instrumentation for the twin 6.5~m Magellan Telescopes, as well as for the Giant Magellan Telescope.  The second is the Smithsonian Astrophysical Observatory (SAO), a public institution that develops instrumentation for both NASA (in the way of orbiting observatories such as the Chandra X-ray Observatory) and for ground-based telescopes such as the MMT 6.5~m Telescope and Giant Magellan Telescope. Both Carnegie Observatories and SAO are large research organizations, each having a subset of personnel dedicated to developing instrumentation for ground-based optical telescopes; Carnegie having approximately a dozen staff, and SAO having about three dozen.  Carnegie Observatories is hard-money funded.  SAO is soft-money funded with staff size fluctuating as needed to meet the workload. 

\subsection{The Department of Energy}

The Department of Energy (DOE) is a somewhat recent entry into the world of astronomy and instrumentation.  Fermilab worked on the development of the original SDSS Telescope, spectroscopic pipeline, and fiber cartridge system, and later worked on the development of the Dark Energy Camera\cite{2015AJ....150..150F} for the Blanco 4~m telescope at the Cerro Tololo Inter-American Observatory (CTIO).  It is not clear if Fermilab is currently involved in any new instrumentation work.  Lawrence Berkeley National Lab has been leading the development of DESI (Dark Energy Spectroscopic Survey)\cite{2019BAAS...51g..57L}.  SLAC National Accelerator Lab and Brookhaven National Lab have been working on the development of the Large Synoptic Survey (LSST) camera. These DOE projects are largely dedicated to the investigation of Dark Energy, hence DOE involvement is natural. What is not clear is whether these groups will continue to be involved in astronomy in the future.  It is also not clear how many personnel at these various institutions are dedicated to astronomy instrumentation.  The DOE has the advantage of what are often very good facilities, and large groups of technical staff with a broad range of expertise, sometimes well matched to the work we do in the astronomy community, and sometimes not.  

However, the matrixed structure of the government labs can be less efficient, and as a consequence, more costly.  Fewer multi-disciplined engineers engaged full-time throughout the life of the project requires fewer meetings, fewer emails, less management, and less systems engineering, and is cheaper than a matrixed organization with many engineers, spread over multiple divisions/institutions, each constrained to their area of expertise, working a shorter duration over the life of the project, requiring a much greater degree of communication and coordination.  

Of course, we should be careful not to be too critical of the government labs when it comes to cost.  Certainly there is a tipping point, beyond which projects are too large and/or complex for small university groups, at which point the size and often superior facilities of the government labs, as well as the permanent staff, are necessary to carry out the work.

\subsection{Financial Support for Groups in the US}
Where the money comes from to fund various instrument groups in the US varies.  

Research institutes like Carnegie Observatories are funded largely by an endowment, with some funding for specific projects coming from, e.g., the National Science Foundation, private donations, or government commitments as in the case of the Giant Magellan Telescope. In general though, staff positions are paid for by endowment and not at risk if the small sources of soft project money go away.

Government labs are funded by the government. And while appropriations can change with time, in general these funds are secure, with little risk of losing key personnel.

For Universities, the situation varies. Some institutions such as UT Austin and OSU are supported largely by hard money in the form of State funds, either directly to the institution, or indirectly through a supporting observatory; although these funds do need State budget approval at some regular cadence.  Others such as JHU and Penn State are entirely soft money with funding coming from either internal research projects or external collaborations. Then there are hybrid situations such as Caltech, UCSC where staff are in principal reliant on soft-money support, but State and institutional funds for supported observatories (i.e. Palomar Observatory, Lick Observatory, Keck Observatory and the Thirty Meter Telescope), naturally lead to a flow of instrument projects that need full-time staff to develop; institutional affiliation serves as a conduit to the group.

\subsection{The Takeaway}
\label{sec:takeaway}

Of the eleven institutions listed in the Table~\ref{tab:census}, which represents almost the entirety of the \emph{traditional} US instrument building community, outside of Arizona/Steward, CfA, and Carnegie, there are only 116.5 FTEs.   How will the US develop, in parallel, three first light facility instruments, and three second light facility instruments, for \emph{two} ELTs, with just over a hundred FTEs when it will take tens of FTEs per instrument per year?   The collective staff at institutions not in this table is not nearly enough to make up the difference, not even close.   And much of the staff at the above-mentioned institutions are, for the foreseeable future, invested in instrument development projects for the current suite of 4~m to 8~m class telescopes.  \emph{In short, the pool of university instrument builders in the United States is not nearly large enough to develop the instruments needed to take advantage of the enormous investments being made in the US ELTs}.  Some resources will come from partner countries and possibly Government labs; but do these sources have the right mix of experienced talent to fill the void?

The other takeaway from Table~\ref{tab:census} is that, of the eleven groups listed, only a few, maybe, are sizeable enough now to build a 30~m facility class instrument.  And that could reasonably be debated.  Most of the groups would need to scale up to accommodate the workload, and some would need to scale up significantly.

Finally, most of the personnel that make up the total in Table~\ref{tab:census} are on soft money, and these positions are only as secure as the likelihood of follow-on projects to replace the one that is currently keeping them employed. Of the total, 67\% are soft money, and if UT Austin were removed from the calculus the soft-money fraction grows to 80\%.

Overall, the situation is far less than ideal.

\section{The Challenges Faced at Johns Hopkins University}
\label{sec:jhu}

To illustrate the challenges faced by university-based instrument groups, we highlight the issues facing the Instrument Development Group at Johns Hopkins, which essentially boil down to three things: funding, people, and vendors.

\subsection{Funding}
Maintaining a continuous flow of funding sufficient to sustain a relatively large soft-money group as we have at JHU is a challenge, especially if continued funding relies on a continuous stream of winning grant proposals.  The IDG has managed to sustain itself, and double in size, over the past twenty years, in large part by collaborating with other institutions.  The group has also benefited from diversifying; i.e. working on space programs, building instruments for condensed matter physics, and for national defense. And the group has benefited from being in close proximity to its sister institution, the Johns Hopkins Applied Physics Lab, as well as to NASA Goddard Space Flight Center, the Space Telescope Science Institute, and the National Institute of Standards and Technology.    The IDG has built instruments for all of them.  It is the funding from these projects, along with funding from JHU-sponsored projects, that has kept the group going.

But the projects are getting larger, requiring larger teams, which require more funding to maintain long-term.  At the same time money available for new instruments is limited, at least when compared to the needs of the community.  There are fewer instruments, requiring more funding, in a funding-limited environment.  It is winner take all; think ALMA, LSST, etc.  And the trend of having a small number of projects absorbing huge fractions of available instrument funding is going to be exacerbated by the development of 30~m telescopes. These are largely private endeavors and the substantial funds to develop these instruments are not likely to go to non-member university groups like the IDG, at least not in a big way.  Full disclosure, the IDG has done a small amount of work for two GMT instruments to date but has no plans for future work at the moment.

The other dimension to the funding problem is having enough money to actually build the instruments. It is the case that virtually all instrument budgets are \emph{success oriented} at the proposal stage (meaning everything goes perfectly throughout the development process; i.e. no surprises). This does not typically happen.  Instrument builders know this,  but to compete for grant funding most proposals adopt the success oriented budget approach, hoping to find a path to additional funds when things do not go as planned. The same can be said of schedules, of course. 

All that said, funding is a challenge in a soft-money world.  For those few instrument groups with hard money staff, these issues and that of survival in an uncertain environment are less of a concern. However, limited funding opportunities constrain how effective and productive the US instrumentation programs can be.

\subsection{Technical Staff}
Good experienced technical people are everything in this business. Trained, motivated, creative, intelligent, skilled, jack-of-all trade people.  They are the difference between an instrument that under-performs and one that works exceptionally well.  An Engineer with these qualities is worth at least two average engineers in productivity, and engineers with these qualities can plug into almost any phase of a project.  You can always find something for these people to do.  There is no downtime.  They are efficient.  The success they create is what brings in future projects.  These are the people we look for and try to nurture.

Finding talented people is a challenge. They are in short supply and in high demand.  The skills needed for astronomical instrumentation are not taught in typical engineering programs. Having the project funding to hire good engineers when you do find them is a particular challenge for instrument groups operating on soft-money.  These people do not just show up at your doorstep when you need them.  The tendency is for them to show up when you cannot afford them, 
or for them to be lured away to industry when funding uncertainty is a concern. It is of particular note that in fixed-duration, grant-funded instrument projects there is an extreme risk of losing key personnel in the crucial latter stages of the development, when they start to see the finite horizon and search for the next job.


\subsection{Key Vendors}
\label{sec:vendors}
An issue faced at JHU, MDO, and other groups as well, is finding quality vendors for niche components and services that are key to the business of building optical instruments for astronomy, such as large-format Volume Phase Holographic (VPH) gratings, multi-layer dielectric coatings on large optics, and detectors or detector controllers. In some cases, there are only one or two vendors producing the product or service required.  Personnel changes or change of management emphasis within a company can lead to dramatic changes in the quality of a product or service that is required. 
Astronomers need high technology products, in large sizes, and in small quantities. None of these attributes lend themselves to a typical profit-centered company structure. The resulting scarcity of firms is exacerbated by mergers and acquisitions, the resultant consolidation into larger companies that are often less interested in our niche.

Take large format VPH gratings, for example.  Kaiser Optical Systems, Inc. (KOSI) is the only company building large format VPH gratings of sufficient quality for the instruments we build.  And while the efficiency of the large format gratings they produce is quite good, the transmitted wavefront error produced by their large-format exposure system is not so good, and is in fact problematic.  But there is nowhere else to go, so we live with it.  We wish this problem could be solved. It should be noted that the large format system, especially the laser, is quite expensive and astronomy gratings do not make enough money to justify the development costs for KOSI. This is a big part of the problem.


There are two other resources for VPH gratings other than KOSI.  WASATCH Photonics, a commercial company, and the Goodman Grating Lab at the University of North Carolina (UNC) run by Chris Clemens.  Both are viable resources.  It has been our experience however, that in general the gratings produced by WASATCH have reduced efficiency compared to KOSI.  And the lab at UNC only produces small format gratings for internal researchers and collaborators, and access to the lab is limited as it is not intended to be a community resource.  If properly funded the UNC lab could become a valuable community resource.
 
Another example is detectors, which are crucial to the success of an instrument and can account for a major portion of the cost. Even CCDs, which have been around for decades are not easily obtained in science quality and the recent take-over of E2V and DALSA by Teledyne has concentrated our sources for these detectors in one company that also has a monopoly on science grade HgCdTe NIR detectors. This concentration leads to very high prices. There are R$\&$D-level CCD developments at certain national labs, but these are not generally accessible to smaller university-based projects.

\section{The Perspective from University of Texas at Austin $/$ McDonald Observatory}
\label{sec:UTAMDO}

Among US instrumentation programs, UT Austin/McDonald Observatory (UTA$/$MDO) has a relatively large group and provides a different perspective to JHU on the challenges of instrument development at the current time. There are common challenges in technology development and skilled technical staff retention, but the funding sources for instrument development are more diverse, coming from Texas State appropriations that support MDO, from grant funds, and from private and foundation philanthropy.

State funds are not guaranteed; they are appropriated every biennium and do fluctuate, but provide a valuable base for the instrument group by covering the majority of the technical group salaries. For the purposes of this White Paper, we can consider these positions as hard money, though they are not tenured. There are a small number of staff on soft money associated with specific projects. Observatory funds and the machine shop are used for smaller instrument projects. Hardware costs for large instrument development projects rely on a combination of grant funding and philanthropic donations. Sometimes startup or Chair funds from faculty cover part of the cost.

UTA/MDO instrument development is focused foremost on local facilities (McDonald Observatory and the upgraded Hobby-Eberly Telescope (HET\cite{HETDEX_instrumentation_2021})). The group has also supported development of instruments for other telescopes (e.g. IGRINS\cite{2014SPIE.9147E..1DP} near infrared high-resolution spectrograph that has been deployed on DCT and Gemini). Often instrument concepts are developed and proven on the smaller telescopes before being applied to the HET or other facilities. A prime example of this is the VIRUS\cite{2006SPIE.6269E..2JH, HETDEX_instrumentation_2021} replicated spectrograph, the prototype\cite{hil08a} for which  was fielded at the MDO 2.7 m Smith Telescope before the 156 channels were constructed for the HET. VIRUS was funded approximately equally by the NSF, Philanthropy, and Texas State$/$partner funds. 

A key component of the instrument program has been training the next generation of instrumentalists, since MDO provides extensive observing facilities and the opportunity to be involved in developing or commissioning instruments. This role is considered a core part of the mission of the program. Students graduating at both the undergraduate and graduate level go on to technical careers in astronomy and industry, and most recently in big data analytics. There are relatively few opportunities in astronomy for our graduates, and postdoctoral positions are often overshadowed by the opportunities and salaries available in the private sector. Providing a sustainable career path as well as opportunities for training is key to maintaining vitality in US instrument development.

The long-term funding for technical staff positions has allowed a depth of experience within the group that would be hard to sustain in a less certain funding environment. The majority of the technical staff have in excess of ten years’ experience designing and deploying forefront instrumentation. In spite of this advantage, technical staff retention is a constant concern, since highly trained individuals with extensive instrument experience in astronomy are in demand by other programs, large telescope projects, and industry.

\section{Discussion}
\label{sec:discussion}

In the United States, instruments have historically been built by small instrument groups, a fraction of the size of our major European counterparts. As instruments have become more complex, this begins to be a problem.  Today very few universities have a team large enough and skilled enough to build instruments for the coming generation of 30~m class telescopes, and most would be challenged to develop an 8~m class instrument.  \emph{This is the core issue we face}.  The workaround has been collaborations amongst universities that, combined, have the depth and requisite skill sets.  This is a necessity, but from a strictly engineering perspective it presents challenges in management, interface control, orphaned subsystems, etc.  It costs more and it takes longer.



If we examine the issue of small, understaffed teams lacking proper facilities, the problem stems directly from the funding model.  All funding, i.e. funding for hardware \emph{and} labor, goes to the lead investigator.  Few lead investigators can put together a string of winning instrument proposals long enough to build a team of sufficient depth and skill level to develop a modern instrument. In fact, none have.  Why?  Because we compete the science, not the technical teams.  And the scientific landscape is very competitive.  It is very hard to win twice in a row.  


Hence, without hard-money or creative thinking, instrument groups in the US do not grow and typically do not survive long-term.  \emph{This is the problem}.  The result is a sparse network of small instrument groups that come and go with the success and failure of the lead investigator.  Those soft-money groups that have survived, have done so as a result of having sufficient technical talent to attract new projects, to be a valued resource when lead investigators lack the technical resources at their home institution, or by diversifying, i.e. working on projects for space observatories, or other areas of physics. This is how the Instrument Development Group at Johns Hopkins University has manged to stay viable. Of course, implicit in this model is that the instrument group not be tied to the scientific interests of a single lead investigator.  At JHU the emphasis has been more on keeping the talent than the particular flavor of science the group supports.

Something should be said about the successes/failures resulting from the soft-money mode of instrument development in the US.  When we look honestly at the track record, the results are mixed.  There have been some tremendous successes; the Sloan Digital Sky Survey\cite{2000AJ....120.1579Y} comes to mind.  There have been failures; Gemini has struggled to develop quality instruments over the years, despite having two excellent telescopes.  An important realization here is that, in all cases, be it success or failure, the science was never the issue. Instruments do not fail to perform because the science was not compelling enough.  
Instruments fail because they do not perform to expectation, or because of a mismatch between science goals and technical implementation, or because they are delivered so late that the competitive landscape has already moved on.
The difference ultimately points, in most cases, to the technical team carrying out the work.  Did the team build an instrument that meets the requirements necessary to achieve the science?  That typically hinges on whether or not the technical team has the skills, knowledge, experience, and motivation to get the job done.  
Success also depends on whether the flow-down of science to technical requirements was realistic, which requires experience and skill.  Ultimately, it's people.

What should we do about this?  We should identify those groups with sufficient expertise to develop large-scale ground-based instruments and, if possible, direct the work to them.  We should try to support the groups that have well-established track records of building quality instruments to help ensure that these groups will be there for future.  With any luck we can grow the size of these groups so we have a ready resource for the next generation of large telescopes that is sure to come, training the next generation of instrument builders along the way and hopefully serving as a place where young instrument builders can be trained and then have a long, stable career.  It is doubtful we could ever achieve what Europe has in the way of permanent, hard-money funded instrument development centers, but we can work to 
maintain and secure the vitality of the University-based approach.

On the technology side, the challenges we face fall into two categories.  First is a dearth of vendors in key technology areas that are vital to our business: detectors, gratings, coatings, etc.  Take for example Teledyne, which now owns e2v and DALSA, and has a near monopoly on the scientific detector market, as noted in Section~\ref{sec:vendors}. 
Fortunately, Teledyne produces quality detectors, but there is little competition, and the announcement recently that DALSA would no longer be accepting commercial orders has nearly put Semiconductor Technology Associates (STA) out of business in the CCD market\footnote{Update: STA has subsequently secured their business model}, leaving Fairchild and Japanese CCD maker Hamamatsu as the only viable competition.  It should be noted that, for a variety of reasons, the majority of visible ground-based instruments use detector systems from either e2v or STA.  It would be a real tragedy if STA left the business, but their business model looks to be in danger. This emphasizes how tenuous some of our key technology is.

A similar story exists with VPH gratings, the technology used in most modern spectrographs having moderate resolution.  As noted in Section~\ref{sec:vendors}, Kaiser Optical Systems, Inc. is a standout amongst only a few VPH grating sources world-wide.  However, their large-format gratings have a key deficiency in the way of transmitted wavefront error, a problem KOSI has not been willing to invest resources in fixing.  
KOSI works with astronomers due to the interest of a small group within the company and a retirement could result in this resource drying up\footnote{Update: Unfortunately, KOSI has subsequently chosen to no longer build custom gratings for astronomy}.
Should the community consider trying to work with KOSI to solve the deficiencies in their exposure system, thereby increasing the value of astronomy to the company, or work with the Goodman Lab at the University of North Carolina to build a large-format exposure system to produce high quality large VPH gratings?  
The UNC effort demonstrates that it is possible to create a center for technology that has the best interests of astronomy at heart, and is immune from commercial pressures.
However, that resource is very dependent on small internal funding and the desire of the Goodman Lab leadership to secure the technology for Astronomy, so it is tenuous at best.

Then there is the business of multi-layer dielectric coatings on large optics; some of which can be quite challenging. World-wide, there are very few vendors capable of producing quality coatings on large-scale optics.  And competition for vacuum chamber time is a real consideration, especially when competing with defense contractors who also rely on these same resources. Is there something we can do to control our destiny? These coatings play a crucial role in the instruments we build and have the potential to drive performance and schedule in some cases.

The second challenge we face in the area of technology is new technology to advance the field.  And this is as much an opportunity as it is a challenge. In fact, the challenge is to create the opportunity. What technologies can we advance now that will push the state of the art in ground-based instrumentation?  And how do we transition these technologies from the lab to the community? Here the US is, at a minimum, on even-footing with the Europeans, but the success of e.g. the on-chip beam combiner of VLTI/GRAVITY at ESO reminds us that competition is strong.  With the vast majority of instrument development taking place at universities in the US, there is no reason we cannot shine here.

\section{How the National Science Foundation Can Help}
\label{sec:NSF_help}

The National Science Foundation can help address the deficiencies discussed above by creating Centers of Excellence in ground-based instrument development.  We can imagine two types: Engineering Centers and Technology Centers.

\subsection{Engineering Centers of Excellence}
\label{sec:IRAD}
Engineering Centers of excellence would be those instrument groups that: a) have a strong track record of building quality astronomical instruments; b) have sufficient full-time technical staff to cover the majority of the skill sets needed to develop a modern instrument; c) have the facilities needed to develop modern instruments, and d) are funded by soft money or a mix of hard and soft. 

To make these Centers successful, two things need to happen.  One, the NSF needs to direct money to the centers in a strategic way.  And two, the NSF needs to direct funding for instrument projects to the centers; somehow.

As for strategic investment, Engineering Centers of Excellence could receive funding for specific needs, such as strategic hiring, proposal preparation, publications, facilities and students. The funding used to hire personnel ensures that the team has the staff it needs to carry out the work \emph{before} it is needed.  Proposal funding helps the team to develop robust instrument concepts and to write solid proposals, which should help produce more robust budgets and schedules for build proposals.  Funding for facilities ensures that groups have the specialized equipment needed to build state-of-the-art instruments.  Student funding trains the next generation of instrument builders and would allow the Centers to engage more with students as a core mission, rather than as a side benefit of a particular specific instrument project.  And funding for publications helps to disseminate the knowledge gained by these expert groups to the community.  The proposal funding is actually a key component of this concept since it provides money to support \emph{any} astronomer's instrument concept, regardless of the institution where the astronomer resides.  It opens up the possibility of being an instrument PI to anyone; this is not the case now.   

Directing instrument funds to the Centers is a bit trickier, as it needs to be done in a fair way.  However, most would consider it fair if the same process in place now to select instruments for funding were used, with one caveat: there needs to be a weighting factor for technical team experience/track-record.  Engineering Centers should receive the highest marks in this regard, giving them an advantage, encouraging them to write proposals in the first place, and encouraging astronomers to use these Centers to develop their instrument concepts; something the Centers could in fact afford to do since the NSF has provided funding exactly for these things.  Think of favorably weighting the proposals that utilize the Engineering Centers as favorably weighting the proposals that target the science ranked most highly by the Decadal Survey.

There are no doubt other ways Engineering Centers could be structured/funded, and individual institutions may have a preference.   But core to this idea is that funding goes to those groups that can best utilize it for the greater good of the US astronomical community.  The ultimate goal is to build instruments of the highest performance, in the shortest amount of time as is reasonable, with budgets that will likely be less than desired.  

\subsection{Technology Centers of Excellence}
\label{sec:tech}

A Technology Center of Excellence would be one that focuses on transitioning technology critical to the astronomical community to the community.  A committee appointed by the NSF would identify key technology areas that are vital to the community and are in need of support. Priority should be given to core technology, which is used now, and for which there are limited sources, or where the current performance is limiting the science that can be realized. Research groups could apply with a specific proposal that addresses the need; in essence, applying to become a Technology Center of Excellence.  Funding would be granted to those groups that are deemed likely to succeed based on a demonstrated success in that field. Technology centers that do succeed will no-doubt get support from Europe, Japan, and other countries that also find them to be a valuable resource, and the cash-flow from these additional sources would lessen the burden on the NSF.   

The grating lab at the University of North Carolina is an example of a possible Technology Center, a venture where NSF funding could pay off.  As it is now, our community is a retirement away from not having a viable path to large-format high throughput volume phase holographic gratings. Investment in the lab at UNC could not only provide a backstop against this real possibility, but it could also address the issue of the poor transmitted wavefront we see in KOSI gratings. And it could potentially provide these benefits at a lower cost to the community.

Another example is the adaptive optics group at the University of Arizona, which is a real bright-spot for the US in a field that has been dominated by the Europeans; the French in particular.  Funding to support the University of Arizona AO lab as a Technology Center could not only advance adaptive optics technology but also increase the US footprint in a field for which there are currently very few players in the US, one that is so critical to the success of thirty-meter class telescopes, which we are investing billions of dollars in. 

There are other potential technology centers that come to mind: surface-relief gratings, detector characterization; detector technology; fiber positioning robots; astrophotonics (fibers and waveguides); digital micromirror device technology for astronomy, and data reduction software to name a few.  These are areas where true expertise is limited to a small subset of the community, yet advances in these areas can potentially have a very large impact on the larger community as a whole.

\section{Conclusion}
\label{sec:conclusion}

This white paper addresses the key challenges faced by the US ground-based telescope instrument building community in the era of extremely large telescopes, issues that must be considered if the US is to have a vibrant future in astronomy and compete head-on with the Europeans who are better funded and have deeper, more stable, pools of talent.  

Above all, the primary issue we face in the US is people; acquiring, nurturing, and retaining, in a funding-limited regime, the specialized talent needed across a vast array of disciplines to build modern instruments.  This is a particular challenge for soft-money funded university instrument groups, which have historically developed the ground-based telescope instruments in the US, and will be relied on for the thirty meter class instruments as well.  At present, the vast majority of these groups lack the full range of talent and facilities needed to build a thirty meter class instrument.   Given the complexity of modern instruments having the right people and facilities is essential for success, and all the money spent on this next generation of telescopes will be in jeopardy if the instruments we build for them fail to perform to expectation.

A second but equally pressing issue is technology, whether it be advancing the state of the art or growing the vendor pool in key technology areas where the community may be limited to a single vendor to supply the product or service needed.

We believe the National Science Foundation should play a key role in working to address these challenges in coordination with the community. As discussed, establishing Centers of Excellence in instrument development is one way to address the problem.  There are no doubt others.  The goal here is to draw attention to the challenges more than it is to solve them. And we look forward to working with the community and the NSF to chart the most productive path forward.

\section{Epilogue}
\label{sec:epilogue}

Since submission of this white paper in January of 2020 the ground-based instrumentation landscape has changed very little.  The number and size of university groups is roughly the same.  

The limited vendor pool in key technology areas like detectors and gratings is still an issue.  In fact since the release of this white paper, Kaiser Optical Systems Inc. has announced that they are no longer accepting orders for custom gratings, as feared might happen. This leaves only one commercial vendor (Wasatch Photonics) and one small university research lab (UNC Goodman Lab) in the US to produce VPH gratings for the astronomical community, an unfortunate reality given that outside of the US there is possibly only one other VPH grating manufacturer worldwide; a small university research group in Brera, Italy.  This is a significant technology risk area for the community.  The development of VPH gratings is as much a craft as it is an engineering process and the learning curve to develop quality gratings is steep, taking many years to develop. Once the knowledge is gone it is very hard to reconstitute in a vacuum; an expert is needed.  The Goodman Lab, which operates as a cost recovery center, is a concrete example of where external investment would benefit the community. The lab currently employs only one  technician and has only one active project.  Once that project is completed, there is real risk that the lab will close, unless a new project materializes or an external source of funding is found.  Clearly there is a need for near-term and sustained investment in this critical technology. 

In the scientific detector industry, Semiconductor Technology Associates is still able to place orders with Teledyne DALSA, for now.  STA has also recently procured a few prototype CCD wafer lots from an alternate foundry.  Both are promising developments in the wake of the buyout of chip maker DALSA by Teledyne, and a relief given that STA has made significant contributions to the ground-based community over the years, as has Teledyne and e2v.  

A significant development since submission of this white paper is the release of the Decadal Survey on Astronomy and Astrophysics 2020 (Astro2020)\cite{NAP26141}.
The survey provides strong support for the US ELT program, recommending \$1.6~B in funding for the Giant Magellan Telescope and the Thirty Meter Telescope, along with support for \$32~M/yr in funding for operations, making the NSF, i.e. the community, a one-third partner in these next generation telescopes.  As discussed here, the instruments for these telescopes will require substantial development effort, and given the current issues faced by the US instrumentation community the commitment to the ELTs makes the current situation a potential crisis. 

This paper serves to shed light on a fundamental problem.  As we enter the era of ELTs, the US community lacks the talent base, facilities, technology resources, and funding to build this next generation of instrumentation, especially considering the parallel need to modernize the existing 4~-~8 m telescope instrument suite.  Certainly instrument funding from the NSF has been on the rise, however the scale of new projects has outpaced the increase in funding, leaving a real financial shortfall when it comes to funding the full complement of instrument programs within the US.  In addition, years of under-funding have led to an ultra-competitive environment where many soft-money instrument groups without sustained funding have become moribund,
depleting the training ground for the next generation of instrument builders and the talent base needed to thrive in the era of ELTs. Lastly, vendor depth in key instrument technologies is in some cases a critical problem. Even under the best of circumstances, resolving these issues will take considerable time, making the need to address them urgent. 
We recommend that the NSF work with community experts to develop a strategic plan to address these issues and build the case for the increase in funding that will no doubt be required.   In particular, NSF leadership will be key in establishing a mechanism for creating and funding centers of technical excellence to secure key technologies that are not tied only to individual instrument projects.

\acknowledgments 
\label{sec:acknowledge}

The authors thank the following individuals for helpful discussions and for providing data on instrument programs in the United States, and internationally:
Ian Bryson - UK/ATC;
Stephen Eickenberry - University of Florida;
Fred Hearty - Penn State;
Mike Lesser - UA/Steward Observatory;
Ian McLean - UCLA;
Eric Persson - Carnegie Observatories (Emeritus);
Rick Pogge - OSU;
Matthew Radovan - UCSC;
Massimo Robberto - STScI/JHU;
Connie Rockosi - UCSC;
Rob Simcoe - MIT;
Matthew Shetrone - UCSC;
Roger Smith - Caltech;
Andy Szentgyorgyi - Harvard-CFA;
Alan Uomoto - Carnegie Observatories;
John Wilson - UVA.

The views expressed in the White Paper are those of the authors, and do not necessarily reflect the views or policies of the institutions at which they are employed.

\bibliography{report} 
\bibliographystyle{spiejour} 

\end{spacing}
\end{document}